\documentclass[twocolumn,prl,showpacs,amsmath,amssymb,aps]{revtex4}

\usepackage{graphicx,latexsym,amsmath,amssymb}
\usepackage{dcolumn}

\newcommand{\beq}{\begin{equation}}
\newcommand{\eeq}{\end{equation}}
\newcommand{\be}{\begin{eqnarray}}
\newcommand{\ee}{\end{eqnarray}}

\def\dd{ \,\mathrm{d} }

\def\+{\dagger}
\def\la{\langle}
\def\ra{\rangle}
\def\<{\langle}
\def\>{\rangle}

\def\atop{\frac{ \alpha_{s}}{8 \pi} G_{\mu \nu}^{a}
 \tilde{G}^{\mu \nu a} }

\newcommand{\exclude}[1]{}
\newcommand{\Lqcd}{\Lambda_{\mathrm{QCD}}}

\begin{document}

\title{The cosmological constant from the QCD Veneziano ghost}

\author{Federico R. Urban and Ariel R. Zhitnitsky}

\affiliation{Department of Physics \& Astronomy, University of British Columbia, Vancouver, B.C. V6T 1Z1, Canada}

\date{\today}

\begin{abstract}
We suggest  that the solution to the cosmological vacuum energy puzzle is linked to the infrared sector of the effective theory of gravity interacting with standard model fields, with QCD fields specifically. We work in the framework of low energy quantum gravity as an effective field theory. 
In particular, we compute the vacuum energy  in terms of QCD parameters and the Hubble constant $H$ such that the vacuum energy is $\epsilon_{vac}  \sim H \cdot  m_q\la\bar{q}q\ra  /m_{\eta'} \sim  (3.6\cdot 10^{-3} \text{eV})^4$, which is amazingly close to the observed value today. The QCD ghost (responsible for the solution of the $U(1)_A$ problem) plays a crucial r\^ole in the computation of the vacuum energy, because the ghost's properties at very large but finite distances slightly deviate (as $\sim H /  \Lqcd $) from their infinite volume Minkowski values. Another important prediction of this framework states that the vacuum energy owes its existence to the asymmetry of the cosmos. Indeed, this effect is a direct consequence of the embedding of our Universe on a non-trivial manifold such as a torus with (slightly) different linear sizes. Such a violation of cosmological isotropy is apparently indeed supported by WMAP, and will be confirmed (or ruled out) by future PLANCK data.
\end{abstract}

\maketitle

{\it \underline{Introduction}} - Fairly thought of as one of the most intricate and puzzling problems in modern physics is that of the present day acceleration of the Universe~\cite{Spergel:2003cb,Riess:1998cb,Perlmutter:1998np}, (see also~\cite{Copeland:2006wr,Sarkar:2007cx} for more up-to-date references).  Observational results tell us that the Universe is permeated with an unknown form of energy density which makes up for about 75\% of the total energy density, which appears to be exactly the critical ratio for which the three-dimensional spatial curvature is zero.  It is customary to associate the ``dark'' energy density with vacuum fluctuations, whose energy density would be proportional to the fourth power of the cutoff scale, associated to the highest energy wave modes, at which the underlying theory breaks down.  If this argument were true, we would be faced with a disagreement between theory and observation varying between 40 to 120 orders of magnitude.   

This letter wants to tackle the problem from an upside down perspective.  More precisely, our guiding philosophy~\cite{Thomas:2009uh} is that gravitation can not be a truly fundamental interaction, but rather it is a low energy effective interaction.  In such a case, the corresponding gravitons should be treated as quasiparticles which do not feel all the microscopic degrees of freedom, but rather are sensitive to the ``relevant excitations'' only. We  note that such a viewpoint represents a standard effective lagrangian approach in all other fields of physics such as condensed matter physics, atomic physics, molecular physics, particle physics.  We should say that this philosophy is neither revolutionary nor new, rather, it has been discussed previously in the literature, see some relatively recent papers~\cite{Bjorken:2001pe,Schutzhold:2002pr,Klinkhamer:2009ri} and references therein.

If we accept the framework of the effective quantum field theory for gravity, than  the basic problem of why the cosmological constant is 120 orders of magnitude smaller than its ``natural'' Plank scale $M_{Pl}^4$ is replaced by a fundamentally different questions: what is the relevant scale which enters the effective theory of gravitation? How does this scale appear in the effective quantum field theory for gravity?  This effective scale obviously has nothing to do with the cutoff ultraviolet (UV) scale $M_{Pl}$; instead, the relevant effective scale must appear as a result of a subtraction at which some infrared (IR) scale enters the physics.

According to this logic, it is quite natural to define the ``renormalised cosmological constant'' to be zero in Minkowski vacuum wherein the Einstein equations are automatically satisfied as the Ricci tensor identically vanishes in flat space (see the discussion after eq.~(\ref{final})). Thus, the energy momentum tensor   in combination with this ``bare cosmological constant'' must also vanish at this specific ``point of normalisation'' to satisfy the Einstein equations.  With this definition the effective QFT of gravity has a predictive power.  In particular, once this procedure is performed, it predicts the behaviour of the system in any non-trivial geometry of the space time. From this definition it is quite obvious that the ``renormalised energy density'' must be proportional to the deviation from Minkowski space time geometry.

This effect can therefore be understood as a Casimir type of vacuum energy.  Notice that the usual Casimir energies (e.g., from photons) are all typically irrelevant in understanding the observed vacuum energy, for they scale as $(L^2d^2)^{-1} \sim H^4$ where $d$ is the distance between plates, $L$ is the size of the plates  and $H$ the Hubble parameter.

{\it\underline{The mechanism}} - With this big picture in mind, it must be clear that the any information related to possible deviations from ${\mathbb R}^4$ flat spacetime could manifest itself in local observables only when there are strictly massless degrees of freedom which can propagate at very large distances $\sim H^{-1}$.  It is well-known that if the corresponding  massless degrees of freedom are free, not interacting fields, they contribute  to the energy momentum tensor as $\langle T_{\mu}^{\mu} \rangle \sim H^4$ as a result of gravitational anomaly~\cite{Birrell:1982ix}. This is an exceedingly small contribution which can be neglected for all practical purposes.

Our crucial observation is that while na\"ively all QCD degrees of freedom can propagate only to very short distances $\Lqcd^{-1}$, there is a unique (unphysical) degree of freedom which is exactly massless and can propagate to arbitrary large distances.  This is the justly celebrated Veneziano ghost~\cite{veneziano}, which is analogous to the Kogut-Susskind (KS) ghost~\cite{Kogut:1974kt} in the Schwinger model~\cite{Schw}. Indeed, despite its being unphysical (and defined with ghost commutation relations) it leads to physically observable consequences: it plays a crucial r\^ole in explaning the value of the $\eta'$ mass as a result of the mixing between the ``would be''  Nambu-Goldstone (NG) boson and the ghost itself.  We note that this mechanism does not violate unitarity as discussed in details in the 2d example~\cite{Kogut:1974kt}.  Nowadays this mechanism  is considered as the standard    Witten-Veneziano resolution of the famous $U(1)_A$ problem~\cite{veneziano,Witten:1979vv}.

The key element of their proposal, which is of chief importance in our discussion, is the topological susceptibility $\chi$ defined as follows,
\be
\label{Q}
\chi\!&\!\equiv\!&\! i\int \!\!\!\dd x   \la 0|T\{Q(x), Q(0)\} |0\ra \, ,\\
Q\!&\!\equiv\!&\!\frac{ \alpha_{s}}{16 \pi} \epsilon^{\mu\nu\rho\sigma} G_{\mu \nu}^{a} G_{\rho\sigma}^{a} \equiv \atop \equiv \partial_{\mu}K^{\mu} \, , \nonumber\\
K^{\mu}\!&\!\equiv\!&\!\frac{g^2}{16\pi^2}\epsilon^{\mu\nu\lambda\sigma}A_{\nu}^a
(\partial_{\lambda}A_{\sigma}^a+\frac{g}{3}f^{abc}A_{\lambda}^bA_{\sigma}^c) \, ,\nonumber
\ee
where $A_{\mu}^{a}$ are the conventional QCD color gluon fields and $Q$ is the topological charge density, and $\alpha_s={g^2}/{4\pi}$.  The standard Witten-Veneziano solution of the $U(1)_A$ problem is based on the assumption (confirmed by numerous lattice computations, see e.g.\ the recent review paper~\cite{Vicari:2008jw} and references therein) that $\chi$ does not vanish despite of the fact that $Q$ is a total derivative $Q=\partial_{\mu}K^{\mu}$. It implies that there is an unphysical pole at zero momentum in the correlation function of $K_{\mu}$, similar to the KS ghost in the Schwinger model~\cite{Kogut:1974kt}.

As we argue below, the existence of this pole is protected, while the corresponding ghost's matrix elements may slightly deviate from their Minkowski values when the theory is defined on a large but finite manifold.  As a result, the topological susceptibility $\chi$ on a finite manifold such as a torus is slightly different from its Minkowski value.  Such a deviation in $\chi$ will be eventually translated into energy density $\epsilon_{vac}(\theta) $ as the topological susceptibility is directly related to it~\cite{Witten:1979vv},
  \beq
  \label{epsilon}
  \chi=-\frac{\partial^2\epsilon_{vac}(\theta)}{\partial \theta^2}|_{\theta=0} \, .
  \eeq
As last step, we recover the $\theta$-dependent portion of the vacuum energy and its deviation from $\mathbb{R}^4$ flat space: this, in our setup, by definition, is the cosmological constant $\rho_{\Lambda}$.

In short, the Veneziano ghost which solves the $U(1)_A$ problem in QCD is also responsible for a slight mismatch in energy density between a finite manifold of size $L\sim H^{-1}$ and Minkowski $\mathbb{R}^4$ space, such that $\rho_{\Lambda}\sim H\Lqcd^3\sim (10^{-3} {\text eV})^4$.  Notice that the appearance of the QCD scale could shed some light on the ``cosmic coincidence'' problem, as it may be the scale at which dark matter forms~\cite{forbes}.

{\it\underline{Dark energy from the Veneziano ghost}} - We begin by reviewing the solution of the  $U(1)_A$ problem as put forward by Veneziano~\cite{veneziano}.  Let us consider gluodynamics without quarks where, at large number of colours $N_c$, the correlation function~(\ref{Q}) is saturated by the Veneziano ghost with matrix elements defined as $ \la 0| K_{\mu}|\mathrm{ghost}\ra=\lambda_{YM}\epsilon_{\mu}$, such that
  \be
  \label{YM}
  \lim_{q\rightarrow 0} i\!\int \!\!\!\dd x e^{iqx}  \la 0|T\{Q(x), Q(0)\} |0\ra =-\lambda_{YM}^2\frac{g^{\mu\nu}q_{\mu}q_{\nu}}{q^2} \, ,
  \ee
where we introduce the ghost's propagator $+i{g^{\mu\nu}}/{q^2}$.  An important remark here is that the propagator has positive sign (it is a ghost), which results in the negative sign in eq.~(\ref{YM}) in contrast with the expected contributions from conventional particles.  This negative sign, as is known, is at the core of the solution of the $U(1)_A$ problem~\cite{veneziano,Witten:1979vv}.  Notice that the correlation function~(\ref{YM}) is sensitive to arbitrary large distances; moreover, the existence of the pole is protected by the topological nature of $Q$, as eq.~(\ref{Q}) requires.  It is because of these properties that, when the system defined in a very large but finite manifold, with typical size $L\sim H^{-1}$, a difference from Minkowski space can arise. We do not make any specific assumptions on the topological nature of the manifold, on whether it is a 4-torus or any other compact manifold, we only assume that there is at least one compact coordinate with size $L\geq H^{-1}$, as observational constraints require~\cite{doug,Starobinsky:1993yx}.

Now we introduce a single light quark with mass $m_q$ (this construction, as is known, can be easily generalised to arbitrary number of  quark flavours~\cite{veneziano}).  We define the corresponding $\eta'$ matrix elements as $ \la 0| K_{\mu}|\eta'\ra=i\frac{\lambda_{\eta'}}{\sqrt{N_c}}q_{\mu}$ such that
   \be
  \label{eta'}
  \lim_{q\rightarrow 0} i\!\int \!\!\!\dd x e^{iqx}  \la 0|T\{Q(x), Q(0)\} |0\ra =~~~~~~~~~~~~~~~~\\ -\lambda_{YM}^2\left[1+\frac{\lambda_{\eta'}^2/N_c}{(q^2-m_0^2)}+... \right]   =-\frac{\lambda_{YM}^2(q^2-m_0^2)}{(q^2-m_0^2-\lambda_{\eta'}^2/N_c)} \, ,  \nonumber
  \ee
where $m_0^2\sim m_q$ is the mass of the ``would be NG particle'' if the ghost's contribution is ignored, while $m_{\eta'}^2= m_0^2+\lambda_{\eta'}^2/N_c$ is the mass of  the physical $\eta'$ field. One can easily check that in the chiral limit $m_q=0$ the topological susceptibility $\chi \sim \lambda_{YM}^2q^2\rightarrow 0$ vanishes as it should for massless quarks. One can also check that the relevant Ward identity (WI) for QCD with light quarks
\be
\label{WI}
\chi\equiv i\!\int \!\!\!\dd x    \la 0|T\{Q(x), Q(0)\} |0\ra = m_q\la\bar{q}q\ra +O(m_q^2) \, ,
\ee
is also satisfied because $m_0^2\sim m_q$.
 
Finally, the famous Witten-Veneziano relation $ 4\lambda_{YM}^2 =  f_{\pi}^2m_{\eta'}^2$ can be obtained by expressing $m_0^2 $ in terms of the chiral condensate, $m_0^2f_{\pi}^2=-4m_q\la\bar{q}q\ra$.  We want to emphasise that the sign in~(\ref{eta'}) remains negative for $q^2=0$, but its absolute value is drastically reduced in comparison with the pure YM case~(\ref{YM}) as a result of the cancellation of the negative contribution from the Veneziano ghost with the positive one from the physical ``would be the NG'' state. The cancellation is exact for $m_q=0$, but a small negative contribution remains for non-vanishing $m_q\neq 0$ as required by the WI~(\ref{WI}).

For our following discussions it is instructive to represent the correlation function~(\ref{eta'}) in coordinate space,
 \be
  \label{x-space}
   i\la T\{Q(x), Q(0)\} \ra= -\lambda_{YM}^2\left[\delta^4(x)-\frac{\lambda_{\eta'}^2 D^c(m_{\eta'}x)}{N_c} \right] \, ,
  \ee
where $D^c(m_{\eta'}x)$ is the Green's function of a free massive particle with standard normalisation $ m_{\eta'}^{2}\int \dd x D^c(m_{\eta'}x) = 1$.  In this expression the $\delta^4(x)$ represents the ghost's contribution while the term proportional to $D^c(m_{\eta'}x)$ represents
the $\eta'$ contribution.  In such a form the correlation function~(\ref{x-space}) is analogous to the known exact result of the 2d Schwinger model, where the corresponding generalisation to a curved and/or topologically non-trivial manifold can be performed and the relevant lessons for the 4d case can be learnt~\cite{UZ}.  We do not assume that equation~(\ref{x-space}) is the exact expression 
for $ i\la T\{Q(x), Q(0)\} \ra $ in QCD; however, we do assume that it grasps all the important features of this correlation function.  In particular, it satisfies the WI~(\ref{WI}). 

Our next step is to study the corresponding Veneziano ghost's contribution to the topological susceptibility and vacuum energy in curved space and on a manifold with a boundary such as  torus, when at least one dimension is large but compact~\cite{doug,Starobinsky:1993yx}.  In what follows we will heavily rely on the results of our accompanying paper on the 2d Schwinger model~\cite{UZ}, where the ghost's contribution to the vacuum energy on a non-trivial manifold can be analytically calculated.

As we mentioned above, a deviation from infinite Minkowski flat space may only occur if a true massless degrees of freedom (which can propagate to arbitrary large distances) does exist in the system.  We do not have any physical massless degrees of freedom in QCD or in the massive Schwinger model (it is the QFT of a single scalar massive particle).  Our main point here that in both cases (QCD and Schwinger model) the corresponding ghost field, despite its being unphysical and unobservable as an asymptotic state, may nevertheless contribute to physically  observable parameters as the solution of the $U(1)_A$ problem (reviewed above) explicitly demonstrates.

The main assumption here is that the correlation function~(\ref{x-space}), defined on a topologically non-trivial manifold such as torus, has this same structure.  Of course some appropriate changes, such as the replacement of the Green's function of free massive particle $D^c(m_{\eta'}x)$ in eq.~(\ref{x-space}) by that defined on a finite on the new manifold, are necessary.  The $\delta^4$ function in eq.~(\ref{x-space}) is defined on $\mathbb{R}^4$ and is also to be replaced by its counterpart defined on the same non-trivial finite manifold.  This argument is supported by the fact that this generalisation saturates the WI with the Veneziano ghost.  What is more important is that this assumption can be explicitly tested in the 2d Schwinger model~\cite{UZ}, where it is shown that the effects of the embedding in compact space can be mimicked by the structure~(\ref{x-space}), which remains thus untouched, with the only difference that the corresponding ghost's matrix elements $ \la 0| K_{\mu}|\mathrm{ghost}\ra'=\lambda_{YM}'\epsilon_{\mu}$ on a finite manifold sightly differ from its Minkowski    values: $(\lambda_{YM}' - \lambda_{YM}) \sim 1/L $.

The key point we are making here is that the corrections due to the very large but finite size $L$ of the manifold are small, $(\lambda_{YM}' - \lambda_{YM}) \sim 1/L \sim H$ but not exponentially small, $\exp (-L)$, as one could na\"ively anticipate for any QFT where all physical degrees of freedom are massive.  We parametrise this departure from flat Minkowski space with a dimensionless coefficient of order one as follows: $\Delta \lambda_{YM} =(\lambda_{YM}' -\lambda_{YM}) = -c (H / m_{\eta'}) \lambda_{YM}$, where $\Delta A=A_{L}-A$ is defined as the difference between the values of $A$ on a large torus and Minkowski space.  The sign of $c$ in general could be positive or negative.  This is our basic ingredeient, and it is well grounded on the explicit computations for the akin 2d Schwinger model~\cite{UZ}.

To be more concrete, as the WI shows~(\ref{WI}) the deviation in the topological susceptibility $\Delta \chi$ is related to that of the chiral condensate $\Delta \la\bar{q}q\ra$.  The corresponding exact 2d computation indeed demonstrates that the magnitude of the chiral condensate on a large torus of size $L$ slightly changes from its infinite Minkowski value as $\Delta \la\bar{q}q\ra= \la\bar{q}q\ra \frac{\pi}{Lm_{\eta'}} ( \frac{1}{|\tau|} - \frac{1}{\tau_0})$, where $\tau = \tau_1 + i\tau_0$ is the Teichm\"uller parameter for the torus.  This formula exhibits the linear term so intensely sought after: despite the fact that, due to the theory containing only a massive degree of freedom, one expects that this correction should be exponentially small, in fact, the modification is linear $(Lm_{\eta'})^{-1}$.  This result comes from the ghost's contribution, which is very sensitive to the specific boundary conditions at very large distances.  We can not perform a similar explicit analytical computation in the 4d case. However, the presence of the Veneziano ghost suggests that the scenario would be very similar to what we observed in the Schwinger model.

Now we want to demonstrate that the small correction we have just obtained is translated into extra energy density when quantising on the torus.  The sign of $\rho_{\Lambda}$ is correlated with the sign of the ghost's contribution to the topological susceptibility.  To be precise, from eqs.~(\ref{epsilon},\ref{eta'},\ref{WI}) we arrive at
   \be
  \label{Delta}
  \Delta\left[\frac{\partial^2\epsilon_{vac}(\theta)}{\partial \theta^2}|_{\theta=0}\right] =-\Delta\chi=
\Delta  \frac{\lambda_{YM}^2m_0^2}{(m_0^2+\lambda_{\eta'}^2/N_c)}\simeq  \nonumber \\
 -c\cdot\frac{2H}{m_{\eta'}}\cdot \frac{\lambda_{YM}^2m_0^2}{(m_0^2+\lambda_{\eta'}^2/N_c)}
\simeq -c\cdot\frac{2H}{m_{\eta'}}\cdot  |m_q\la\bar{q}q\ra  |<0 \, , 
  \ee

The $\theta$-dependent portion of vacuum energy at $\theta\ll 1$ is well known, and for $N_f$ quarks with equal masses is given by $\epsilon_{vac}(\theta)=-N_f |m_q\la\bar{q}q\ra  |\cos(\theta/N_f)$~\cite{Witten:1980sp,Di Vecchia:1980ve}, such that $\partial^2_\theta \epsilon_{vac} = - \epsilon_{vac}/N_f$ (see also the generalisation for finite $N_c$ and unequal $N_f$ masses  in~\cite{halp}).  Therefore, our relation~(\ref{Delta}) for $N_f=1$ can be written in the following final form,
  \be
  \label{final}
  \rho_{\Lambda}\equiv   \Delta\epsilon_{vac}  = 
  c\cdot\frac{2H}{m_{\eta'}}\cdot  |m_q\la\bar{q}q\ra  | \sim c(3.6\cdot 10^{-3} \text{eV})^4 \, , 
  \ee
to be compared with the observational value $ \rho_{\Lambda}= (2.3\cdot 10^{-3} \text{eV})^4$.  The similarity in magnitude between these two values is very encouraging.  It is also important to notice that the non-vanishing result for $ \rho_{\Lambda}$ is parametrically proportional to $m_q$, and only occurs if the $\theta$-dependence is non-trivial.  In particular, in the chiral limit $m_q=0$ when all $\theta$-dependence is gone from every physical observable, the effect under consideration~(\ref{final}) also identically vanishes.  The same phenomenon can be explicitly seen in 2d  massless Schwinger model~\cite{UZ}.  Such a feature can be easily understood if one recalls that the starting point of the calculation is based on the ghost's contribution to the correlation functions for the operators $K^{\mu}$ and $Q=\partial_{\mu}K^{\mu}$ which are intimately coupled to $\theta$, the ghost being related to this gapless excitation (it represents its extra generalised degree of freedom).

Let us also point out that the result~(\ref{final}) is based on our understanding of the ghost's dynamics: it can be analytically computed in the 2d Schwinger model and hopefully it can be tested in 4d QCD using lattice QCD computations.  More than that, the effective Lagrangian descriging 4d QCD chiral dynamics turns out to be exactly the same as that of 2d QED~\cite{dyn}.  Finally, this contribution to the vacuum energy is computed using QFT techniques in a static non-expanding universe; as it stands, it can not be used for studying its evolution with the expansion of the universe (in order to do so one needs to know the dynamics of the ghost field coupled to gravity on a finite manifold~\cite{dyn}).

A final comment on our definition (or prescription) for the physical vacuum energy.  As we have discussed  in the introduction, we define the observable vacuum energy as the differential stress tensor between infinite Minkowski and finite compact spacetime.  Therefore, with this prescription, all the usual contributions such as gluon condensates, or the condensate from the Higgs field, etc., will cancel out in the subtraction as they appear  with almost equal magnitude in both compact size $L$ and non-compact manifolds. The relevant difference will behave as $\exp(-m L)$ due to their massiveness and can be safely neglected.  The Veneziano ghost's contribution is unique in all respects: its masslessness is protected and is therefore the only field linearly sensible to the global topology.

{\it \underline{Conclusion}} - The main result of the present study is that the QCD ghost, which solves the $U(1)_A$ problem, contributes a non-standard Casimir-type term to the vacuum energy of the confined phase of 4d QCD.  This effect is interesting in itself and deserves to be looked for in dedicated numerical approaches.  We therefore urge the lattice community to undertake searches for a topological susceptibility which is sensitive to the boundary conditions, with deviation from its asymptotic value decaying only as $1/L$ in spite of the fact that all physical degrees of freedom in QCD are massive.

We do not claim that the ghost field becomes a propagating degree of freedom capable to produce a Casimir effect.  Rather, we argue that the topological properties of 2d QED and 4d QCD are very similar as both theories are described by the same effective Kogut-Susskind Lagrangian.  In addition, they also both support the construction of the $\theta$ vacuum state resulting from large gauge transformations.  The description in terms of the ghost is a convenient way to account for such kind of physics hidden in the non-trivial boundary conditions.

There is an immediate application to the cosmological constant problem, which the Veneziano pole would be essentially a source for.  This is due to the fact that the ghost's properties at very large distances on a topologically non-trivial manifold slightly differ from its Minkowski ones as $(H/ \Lqcd) $.  Explicit and exact analytic results in the 2d Schwinger model~\cite{UZ} support our claim.  Essentially we claim that the dark energy $\rho_{\Lambda}$ in principle can be studied by doing numerical lattice QCD computations by analysing the $L$ dependence of the topological susceptibility.

Another important consequence of this proposal is the observation that if the cosmological constant $\rho_{\Lambda}$ indeed arises from the finiteness of the torus we live in, than the corresponding topological structure on the scale $1/L\sim H$ can be already probed using the CMB~\cite{doug,Starobinsky:1993yx}.  Our original additional statement here is that dark energy and the topological structure of the Universe are intimately linked one another, and we predict that there must be an asymmetry in the CMB if the cosmological constant $\rho_{\Lambda}$ is   explained by the mechanism suggested in this paper where a preferred direction is determined by the position and the structure of a compact manifold with typical size $L\sim H^{-1}$.  In fact, WMAP data apparently has been pointing towards such kind of asymmetry for quite some time, see the recent paper~\cite{Hansen:2008ym} and references therein.  Hopefully, PLANCK will finally settle the issue in the nearest future, and we have some specific suggestions on how this proposal can be tested~\cite{Urban:2009ke}. 

{\it \underline{Acknowledgments}} - We thank P.~Naselsky for discussions on the observed asymmetry in the CMB, D.~Scott for valuable conversations, A.~Starobinsky for explaining his model~\cite{Starobinsky:1993yx}, and G.~ Volovik for correspondence.  This research was supported in part by the Natural Sciences and Engineering Research Council of Canada.

\end{document}